\documentclass[preprint,showkeys,showpacs,preprintnumbers,amsmath,amssymb]{revtex4}
\usepackage{graphicx}
\usepackage{dcolumn}	% Align table columns on decimal point
\usepackage{bm}	       % bold math

\usepackage{amssymb, amsmath}
\usepackage{mathrsfs}  % required for $\lap$ := $\mathscr L$ ...
\usepackage{color}
\newcommand{\highlight}[1]{\colorbox{yellow}{#1}}
\newcommand{\D}{\partial}
 \newcommand{\ve}[1]{{\bf #1}}
 \newcommand{\vve}[1]{{\bf { #1}}}
 \newcommand{\veg}[1]{{\boldsymbol {#1}}}	% Greek vector 
 \newcommand{\vveg}[1]{{\boldsymbol {#1}}}	% Greek matrix

 \newcommand{\const} {{\rm const}}

 \newcommand{\delt} {{\Delta t}}

\newcommand{\mat}[4]{ \bracket{\begin{array}{ll}
				#1  &#2\\
				#3  &#4
				\end{array}}}

\newcommand{\matt}[9]{ \bracket{\begin{array}{lll}
				#1  &#2  &#3\\
				#4  &#5  &#6\\
				#7  &#8  &#9
				\end{array}}}
\newcommand{\vectt}[3]{ \bracket{\begin{array}{l}
				#1 \\
				#2 \\
				#3
				\end{array}}}

\newcommand{\dt}[1]{\frac {d#1} {d t}}
\newcommand{\abs}[1]{\left|#1\right|}
\newcommand{\bracket}[1]{\left[#1\right]}

\newcommand{\parenth}[1]{\left(#1\right)}

\DeclareMathSymbol{\R}{\mathbin}{AMSb}{"52}

 \newcommand{\excl}[1]{{\backslash\hspace{-0.3em} #1}}

 \newcommand{\ENSO}{{El~Ni\~no} }

 \newcommand{\DI}[1]{\frac {\D #1} {\D x_1}}
 
 \newcommand{\DIDI}[1]{\frac {\D^2 #1} {\D x_1^2}}

\begin{document}

   \title{\bf 
    Normalizing the causality between time series}

        \author{X. San Liang}
	% \footnote {URL: http://people.deas.harvard.edu/$\sim$san/ }}

	\email{sanliang@courant.nyu.edu}
	\affiliation{Nanjing University of Information Science and
	Technology (Nanjing Institute of Meteorology), Nanjing 210044, and\\
	China Institute for Advanced Study,\\
	Central University of Finance and Economics,
	Beijing 100081, China}

         \date{\today}

\begin{abstract}
{
   Recently, a rigorous yet concise formula has been derived to evaluate
   the information flow, and hence the causality in a quantitative sense, 
   between time series. To assess the importance of a resulting causality,
   it needs to be normalized.
The normalization is achieved through distinguishing three types 
of fundamental mechanisms that govern the marginal 
entropy change of the flow recipient.
A normalized or relative flow measures its importance relative to 
other mechanisms.
In analyzing realistic series, both absolute and relative information flows 
need to be taken into account, since the normalizers for a pair
of reverse flows belong to two different entropy balances;
it is quite normal that two identical flows may differ a lot
in relative importance in their respective balances.
We have reproduced these results with several autoregressive models. 
We have also shown applications to a climate change problem 
and a financial analysis problem. 
For the former, reconfirmed is the role of the Indian Ocean Dipole 
as an uncertainty source to the \ENSO prediction. This might partly 
account for the unpredictability of certain aspects of El Ni\~no
that has led to the recent portentous but spurious forecasts 
of the 2014 ``Monster El~Ni\~no''.
	%
	% portentous forecasts of the 2014 ``monster \ENSO'', 
	% which, however, turned out to be computer artifacts.
	%
For the latter, an unusually strong one-way causality has been 
identified from IBM (International Business Machines Corporation) 
to GE (General Electric Company) 
in their early era, revealing to us an old story,
which has almost gone to oblivion, about ``Seven Dwarfs'' competing a giant 
for the mainframe computer market.
%
% for mainframe manufacturing.
% for mainframe computer manufacturing.
%

}
\end{abstract}



\maketitle

\section{Introduction}	\label{sect:intro}

Information flow, or information transfer as it may be referred to in the
literature, has long been recognized as the appropriate measure
of causality between dynamical events\cite{causalmeasure}. 
It possesses the needed asymmetry or directionalism for a
cause-effect relation, and, moreover, provides a quantitative
characterization of the otherwise statistical test, e.g., 
the Granger causality test\cite{Granger}.
For this reason, the past decades have seen a surge of interest in this
arena of research. Measures of information flow proposed thus far include,
for example, the time-delayed mutual information\cite{Swinney},
transfer entropy\cite{Schreiber}, momentary information
transfer\cite{Runge}, causation entropy\cite{JieSun}, etc., 
among which transfer entropy has been proved to
be equivalent to Granger causality up to a factor 2 for linear 
systems\cite{Barnett}.

%
% This has particularly clear when the widely used Granger causality test
% and the concept of transfer entropy proposed
% by Schreiber to quantify information flow\cite{Schreiber} are shown to be
% equivalent up to a factor 2\cite{Barnett} in a linear system. 
% However, it is also known that transfer entropy is biased, 
% depending on the autodependency coefficient in
% a dynamical system; as a result, it may result in spurious
% causalities\cite{transfer_entropy_problems}.
%

Recently, Liang and Kleeman find that the notion of information flow 
actually can be put on a rigorous footing within a given deterministic 
system\cite{LK05}. 
%
% A remarkably concise formula is rigorously derived which
% faithfully and quantitatively reveals the cause-effect relation between
% time series. With it, we show how some outstanding touchstone problems,
% including one raised in a recent PRL paper (Hahs \& Pethel, PRL 107, 12870), 
% can be easily solved. 
%
The basic idea can be best illustrated with a system of two
components, say, $X_1$ and $X_2$. The problem here essentially deals
with how the marginal entropies of $X_1$ and $X_2$, written respectively 
as $H_1$ and $H_2$, evolve.  Take $H_1$ for an example.
Its evolution could be due to $X_1$ its own and/or caused by $X_2$.
That is to say, $dH_1/dt$ can be split exclusively into two parts:
	\begin{eqnarray*}
	\dt {H_1} = \dt {H_1^*} + T_{2\to1},
	\end{eqnarray*}
if we write the contribution from the former mechanism 
as $dH^*/dt$ and that from the latter as $T_{2\to1}$. 
This $T_{2\to1}$ is the very time rate of information flowing 
from $X_2$ to $X_1$. 

To find the information flow $T_{2\to1}$, it suffices to find
$dH_1^*/dt$, since, for each deterministic system, there is a Liouville
equation for the density of the state and, accordingly, $dH_1/dt$ can 
be obtained. In Ref.~\cite{LK05}, $dH_1^*/dt$ is acquired through an
intuitive argument based on an entropy evolutionary law established
therein. The same result is later on rigorously proved; see \cite{Liang13}
for a review. For stochastic systems which we will be considering 
in this study, 
the trick ceases to work, but in \cite{Liang08} Liang manages to
circumvent the difficulty and find the result, which we will be briefly
reviewing in the following.

Consider a two-dimensional (2D) stochastic system
	\begin{eqnarray}	\label{eq:sde}
	d\ve X = \ve F(\ve X, t) dt + \vve B(\ve X, t) d\ve W,
	\end{eqnarray}
where $\ve F = (F_1, F_2)^T$ 
is the vector of drift coefficients 
(differentiable vector field), 
	$\vve B = \mat {b_{11}} {b_{12}}
		       {b_{21}} {b_{22}}$ 
the matrix of stochastic perturbation coefficients, 
and $\ve W$ a 2D standard Wiener process. Let
	$g_{ij} = \sum_k b_{ik} b_{jk}$, and
	$\rho_i$ be the marginal probability density function of $x_i$.
Liang (2008)\cite{Liang08} proves that the time rate of information flowing
from $X_2$ to $X_1$ is
	\begin{eqnarray}	\label{eq:T21}
	T_{2\to1} = -E\parenth{\frac 1 {\rho_1} \DI {F_1\rho_1}}
	  + \frac12 E\parenth{\frac 1 {\rho_1} \DIDI {g_{11} \rho_1}},
	\end{eqnarray}
where $E$ signifies the operator mathematical expectation. This measure
of information flow is asymmetric between the two parties and, particularly, if
the process underlying $X_1$ does not depend on $X_2$, then the resulting
causality from $X_2$ to $X_1$ vanishes.  This is the so-called 
{\it property of causality}, which asserts, in the above language,  
that $T_{2\to1}$ vanishes if $F_1$ and $g_{11}$ are independent of $x_2$.
When $T_{2\to1}$ is nonzero, it may take positive or negative values. 
A positive $T_{2\to1}$ means $X_2$ causes $X_1$ to be more uncertain, while
a negative $T_{2\to1}$ reduces the entropy of $X_1$, and hence functions to
stabilize the latter. For more details, referred to Ref.~\cite{Liang13}.

The above theorem is recently applied to time series analysis. 
Under the assumption of a linear model with additive noise, Liang\cite{Liang14} 
shows that the maximal likelihood estimate of the information flow
in (\ref{eq:T21}) turns out to be very tight in form, involving only the
common statistics namely sample covariances.
%
% that the cause-effect relation
% in terms of Liang-Kleeman information flow between time series 
% can be faithfully estimated from a rigorously established 
% theorem within the framework of differential dynamical systems.
%
Take two series $X_1$ and $X_2$ for example. 
The rate of information flowing
(units: nats per unit time) from $X_2$ to $X_1$ is shown to be
	\begin{eqnarray}	\label{eq:T21_est}
	T_{2\to1} = \frac {C_{11}C_{12}C_{2,d1} - C_{12}^2C_{1,d1}} 
			  {C_{11}^2 C_{22} - C_{11} C_{12}^2},
	\end{eqnarray}
where $C_{ij}$ is the sample covariance between $X_i$ and $X_j$,
and $C_{i,dj}$ is that between $X_i$ and $\dot X_j$, $\dot X_j$ being the
difference approximation of $dX_j/dt$ using the Euler forward scheme:
	\begin{eqnarray}
	\dot X_j(n) = \frac {X_j(n+k) - X_j(n)} {k\delt}.
	\end{eqnarray}
Here $k$ is usually 1, but for highly chaotic and densely sampled series,
$k=2$ should be chosen to avoid getting spuriously large $\dot X_j$ 
due to possible shock structures that make the differencing highly sensitive 
to the error in $X_j$. This formula involves only 
sample covariances, and is hence very convenient to evaluate. 
In addition, it is easy to see that
if $C_{12}=0$ then $T_{2\to1}=0$, but when $T_{2\to1}=0$ the correlation
$C_{12}$ does not need to vanish. 
That is to say, contrapositively, 
{\it causation implies correlation, but correlation does not imply
causation}.
In an explicitly quantitative way, this corollary 
resolves the long-standing debate over causation versus correlation.

	\begin{figure}[h]
	\begin{center}
	\includegraphics[angle=0, width=0.75\textwidth]
		{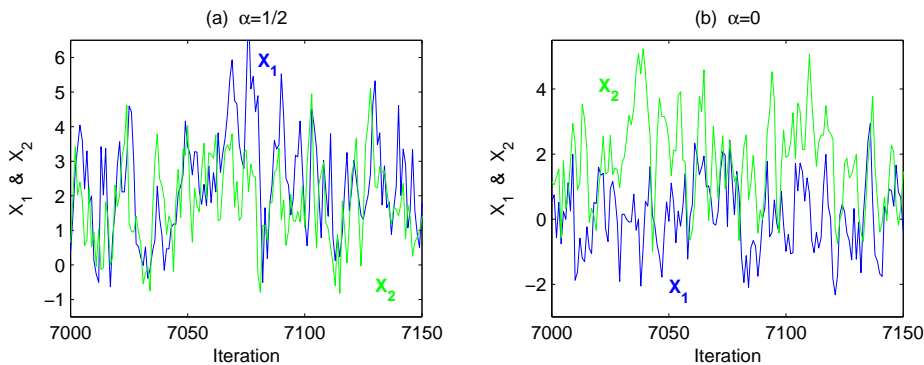}
	\caption{(Color online) 
		A sample path ($X_1$ and $X_2$) of the autoregressive
		process (\ref{eq:ar1}) for (a) $\alpha=0.5$, $\beta=0$,
		and (b) $\alpha=0,$ $\beta=0$.
	 \protect{\label{fig:ar1}}}
	\end{center}
	\end{figure}

In general, the rates of information flow differ from case to case. 
One would like to normalize them in real applications, just as that with
correlation coefficients, in order to assess the importance of the flow
identified, if any. For example, consider two series generated from 
two autoregressive processes:
	\begin{subequations} \label{eq:ar1}
	\begin{eqnarray}	
	&& X_1(n+1) = 0.1 + 0.5 X_1(n) + \alpha X_2(n) + e_1(n),\\
	&& X_2(n+1) = 0.7 + \beta X_1(n) + 0.6 X_2(n) + e_2(n),
	\end{eqnarray}
	\end{subequations}
where the errors $e_1\sim N(0,1)$ and $e_2\sim N(0,1)$ are independent.
Generate a pair of series with 80000 steps on MATLAB, and perform the causality
analysis (with $k=1$), one obtains
  (units are in nats per iteration; same below in this section)
%
%	\begin{eqnarray*}
%	&& \alpha = 0.5, \ \beta=0: \\
%	&&	\qquad\qquad T_{2\to1} = 0.1481 \pm 0.0015,  \\
%	&&	\qquad\qquad T_{1\to2} = -0.0002 \pm 0.0015, \\
%	&& \alpha = 0, \ \beta=0:				\\
%	&&	\qquad\qquad T_{2\to1} = -1.37\times 10^{-5} 
%					 \pm 2.19 \times 10^{-5},  \\
%	&&	\qquad\qquad T_{1\to2} = 1.28 \times 10^{-5}
%					\pm 2.19 \times 10^{-5}.
%	\end{eqnarray*}
%
	\begin{itemize}
	\item[(1)] for $(\alpha,\beta) = (0.5,0)$, \
                  $T_{2\to1} = 0.1481$, $T_{1\to2} = -0.0002$;
	\item[(2)] for $(\alpha,\beta) = (0,0)$: \ \ \ 
                  $T_{2\to1} = -1.37\times 10^{-5}$, 
		  $T_{1\to2} = 1.28 \times 10^{-5}$.
	\end{itemize}
(The results may differ slightly with different series due to the random
number generation.)
For the case $\alpha=0.5$, $|T_{2\to1}/T_{1\to2}| > 740$, one may then
conclude that this is a one-way causality from $X_2$ to $X_1$, as is
indeed true. For the latter, however, one actually cannot say much from the
numbers. Although they are small, they tell  no more than that the
information flows in both directions are of equal importance.
From this particular example, one sees that, though theoretically
the information flows in both directions should be precisely zero, 
in reality the rates evaluated from two time series, albeit very small, 
generally do not precisely vanish.
One then cannot tell whether the causality indeed exists. 
We need to normalize the obtained information flow for one to see
the relative magnitude.

Of course, one may perform statistical test for the results. In the first
example, at a 90\% significance level, 
                 $T_{2\to1} = 0.1481 \pm 0.0015,$ 
		 $T_{1\to2} = -0.0002 \pm 0.0015,$
so $T_{1\to2}$ is insignificant. For the second example,
	$T_{2\to1} = -1.37\times 10^{-5} \pm 2.19 \times 10^{-5},$
	$T_{1\to2} = 1.28 \times 10^{-5} \pm 2.19 \times 10^{-5}.$
That is to say, at a 90\% level, these flow rates are
not significantly different from zero. In this sense, it seems that 
we could indeed infer rather accurately the true causality. 
However, a statistical test just tells how precise the estimate is; it does
not tell how the information flow may weigh in the entropy balance of the
series. Besides, it depends on the length of the series which is irrelevant to 
the parameter to be estimated. To see this more clearly, 
consider the following case:
	$$\alpha = 0.01, \qquad \beta = 0.01.$$
Obviously, the information flows, albeit existing, make only tiny
contributions to their respective series, as the coupling coefficients are
over an order smaller; in classical perturbation analysis, they can be
dropped to the first order approximation. The computed information flows,
at a 90\% significance level, are 
	$T_{2\to1} = (1.648 \pm 0.692) \times 10^{-4},$
	$T_{1\to2} = (1.123 \pm 0.639) \times 10^{-4}.$
These results indicate that they are significantly different from zero---This 
from one aspect testifies to the success of the formalism. However, the small
numbers cannot tell how important they are, since, with a slowly varying 
series, even the dominant flow could be very small. On the other hand, if we
cut the series by half and pick the first 40000 points for the analysis,
then the result will be 
	$T_{2\to1} = (0.653 \pm 0.751) \times 10^{-4},$
	$T_{1\to2} = (1.240 \pm 0.690) \times 10^{-4}.$
So one finds that $T_{2\to1}$ is insignificant while $T_{1\to2}$ is.
(Again, these small numbers may show large fluctuation if series of
different lengths are used, because in reality they are insignificant.)
Can one thus conclude that there is a one-way causality, or can he/she thus
asserts that this shortened series yields a more reliable estimation?
Surely this is absurd. The problem here is that we do need a normalized
flow to evaluate its importance relative to other factors.

% The normalization may not be as simple as it seems to be. A natural 
% normalizer that comes to mind, at the hint of correlation coefficient, 
% might be the information of a series transferred from itself. 
% A snag is, however, that this quantity may turn out to be zero, 
% just as that in the H\'enon map, a benchmark problem we have 
% examined before (see the references in \cite{Liang13}). 
%
% Another snag is, the two way causality actually cannot be normalized 
% together, as that in correlation analysis based on the Cauchy-Schwarz 
% inequality. That is to say, two same rates of information flow may have 
% different relative importance to their respective series.
%
% Since in this framework the information flow from $X_2$ to $X_1$ 
% is the contribution of $X_2$ that makes to the time rate of change of the
% margial entropy of $X_1$, written $H_1$, one may ask whether the rate of 
% marginal entropy change $dH_1/dt$ can be the normalizer. There is a third
% snag, however. As information flow can be positive or negative, $dH_1/dt$
% may be smaller than the former.
%

The normalization is by no means as simple as it appears
with correlation analysis, which is in an inner product
form based on the Cauchy-Schwarz inequality. 
In the following we will see that we need to get down to the 
fundamentals of information flow before arriving at a logically 
and physically sound normalizer.
As what we did before in \cite{Liang14}, this normalizer is estimated 
with the method of maximal likelihood estimation, 
using the given time series (section~\ref{sect:estimation}).
The resulting formula is then validated (section~\ref{sect:validation}) 
with the autoregression example as shown above. 
To demonstrate its diverse applicability, presented subsequently are 
two real world examples, one from 
climate science (section~\ref{sect:climate}), 
another from financial economics (section~\ref{sect:finance}). 
This study is summarized in section~\ref{sect:summary}.

\section{Information flow normalization}
As mentioned in the introduction,
the normalization is not as simple as it seems to be. A natural 
normalizer that comes to mind, at the hint of correlation coefficient, 
might be the information of a series transferred from itself. 
A snag is, however, that this quantity may turn out to be zero, 
just as that in the H\'enon map, a benchmark problem we have 
examined before (see the references in \cite{Liang13}). 

Another snag is, the two way causality actually cannot be normalized 
together, as that in correlation analysis based on the Cauchy-Schwarz 
inequality. That is to say, two information flows of equal size may have 
different relative importances in their respective series.

Since in our framework the information flow from $X_2$ to $X_1$ 
is the contribution of $X_2$ that makes to the time rate of change of the
marginal entropy of $X_1$, written $H_1$, one may ask whether the rate of 
marginal entropy change $dH_1/dt$ can be the normalizer. This might be
appealing, but there is a third snag. As information flow can be positive 
or negative, $dH_1/dt$ may turn out to be smaller than the flow in absolute
value---The so-obtained relative flow would exceed 100\%, a case which we
do not want to see.

   % As elaborated in the introduction, the normalization cannot be a
   % generalization of that for correlation analysis, which is in an inner
   % product form based on the Cauchy-Schwarz inequality. 
   %
All the above tells that information flow normalization is by no means a
trivial task. We need to get to the basics and analyze
how an information flow within a system is derived. 
By Ref.~\cite{Liang08}, the time
rate of change of the marginal entropy of $X_1$ is
	\begin{eqnarray}	\label{eq:dH1}
	\dt {H_1} = - E\parenth{F_1 \DI {\log\rho_1}} 
		    - \frac12 E\parenth{g_{11}\DIDI {\log\rho_1}}.
	\end{eqnarray}
It is actually a result of two mutually exclusive mechanisms: the first
is the information flow $T_{2\to1}$ as shown in (\ref{eq:T21}); the second
is the complement, i.e., the rate of entropy increase without taking
into account of the effect of $X_2$. Denote this latter as $\dt
{H_{1\excl2}}$, Liang (2008)\cite{Liang08} has proved that
	\begin{eqnarray}
	\dt {H_{1\excl2}} = E\parenth{\DI {F_1}}
		- \frac12 E\parenth{g_{11} \DIDI {\log\rho_1}}
		- \frac12 E\parenth{\frac1 {\rho_1} \DIDI {g_{11}\rho_1}}.
	\end{eqnarray}
The right hand side has three terms. The first term is precisely the time
rate of change of $H_1$ due to $X_1$ itself in the absence of
stochasticity. This is the starting point which we have shown in 
2005\cite{LK05} in establishing the rigorous formalism and proved later on
(cf.~\cite{Liang13}). Hence through a careful analysis, the 
increase in the marginal entropy $H_1$ is decomposed into three parts:
	\begin{eqnarray}
	&& \dt{H_1^*} = E\parenth{\DI {F_1}}, 		\label{eq:dH1star}\\
	&& \dt{H_1^{noise}} = 
		- \frac12 E\parenth{g_{11} \DIDI {\log\rho_1}}
		- \frac12 E\parenth{\frac1 {\rho_1} \DIDI {g_{11}\rho_1}},
				\label{eq:dH1noise}
	\end{eqnarray}
as well as $T_{2\to1}$ in (\ref{eq:T21}),
which correspond to, respectively, the contribution due to $X_1$ itself, 
the stochastic effect, and the information flowing from $X_2$.
Note this decomposition does not appear explicitly in the marginal entropy
evolution equation~(\ref{eq:dH1}), as the two stochastic terms cancel out. 

The normalization is now made easy. Let
	\begin{eqnarray}	\label{eq:normalizer}
	          Z      \equiv \abs{T_{2\to1}} + 
				\abs{\dt {H_1^*}} +
				\abs{\dt {H_1^{noise}}}.
	\end{eqnarray}
Obviously it is no less than $T_{2\to1}$ in magnitude, and cannot be zero
unless $X_1$ does not change, a situation that is excluded 
in time series analysis. We may therefore pick $Z$ as the normalizer, and
define
	\begin{eqnarray}
	\tau_{2\to1} = T_{2\to1} / Z.
	\end{eqnarray}
This way if $\tau_{2\to1} = 1$, the variation of $H_1$ is 100\% due to the
information flow from $X_2$; if $\tau_{2\to1}$ is approximately zero, $X_2$
is not causal. Therefore, $\tau_{2\to1}$ assesses the importance
of the influence of $X_2$ to $X_1$ relative to other processes.

It should be pointed out that, the above normalizer applies to $T_{2\to1}$
only. For $T_{1\to2}$, it is 
         $$\abs{T_{1\to2}} + \abs{\dt {H_2^*}} + \abs{\dt {H_2^{noise}}},$$
which may be quite different in value.
This from another aspect reflects 
the asymmetry between $T_{2\to1}$ and $T_{1\to2}$.

\section{Estimation}	\label{sect:estimation}
As in Ref.~\cite{Liang14}, consider a linear version of the 
stochastic differential equation (SDE) (\ref{eq:sde})
	\begin{eqnarray} 	\label{eq:gov}
	d\ve X = \ve f + \vve A \ve X dt + \vve B d\ve W,
	\end{eqnarray}
where $\ve f$ is a constant vector, and
$\vve A = (a_{ij})$ and 
$\vve B = (b_{ij})$ are constant matrices. Initially if $\ve X$ obeys a
Gaussian distribution, then it is a Gaussian for ever, i.e.,
	\begin{eqnarray}
	\rho(\ve x) = \frac 1 {2\pi \sqrt{\det\vveg\Sigma}}
		e^{-\frac12 (\ve x - \veg\mu)^T \vveg\Sigma^{-1} 
			    (\ve x - \veg\mu)},
	\end{eqnarray}
with the mean
$\veg\mu$ and covariance matrix $\vveg\Sigma$ governed by equations
	\begin{eqnarray}
	&& \dt{\veg\mu} = \ve f + \vve A \veg\mu, 	\label{eq:mu} \\
	&& \dt{\vveg\Sigma} = \vve A \vveg\Sigma + \vveg\Sigma \vve A^T 
			+ \vve B \vve B^T.	\label{eq:sigma}
	\end{eqnarray}
So Eqs.~(\ref{eq:dH1star}) and (\ref{eq:dH1noise}) can be explicitly evaluated:
	\begin{eqnarray}	\label{eq:dH1star_lin}
	\dt {H_1^*} = E\parenth{a_{11}} = a_{11},
	\end{eqnarray}
and
	\begin{eqnarray*}
	\dt {H_1^{noise}} 
	&=& -\frac 12 E\parenth{g_{11} \DIDI{\log\rho_1}}
	   -\frac 12 E\parenth{\frac1{\rho_1} \DIDI{g_{11}\rho_1}}  \cr
	&=& -\frac12 E\bracket{g_{11} \parenth{-\frac 1 {\sigma_1^2}}}
	  -\frac12 \iint_{\R^2} \rho_{2|1} \DIDI {g_{11}\rho_1} dx_1dx_2 \cr
	&=& \frac12 \frac {g_{11}} {\sigma_1^2}
	  - \frac12 \int_\R \DIDI {g_{11}\rho_1}
		\parenth{\int_\R \rho_{2|1}(x_2 | x_1) dx_2} dx_1,
	\end{eqnarray*}
since neither $g_{11}$ nor $\rho_1$ depends on $x_2$.
But $\int_\R \rho_{2|1} = 1$, and $\rho_1$ is compactly supported, the
whole second term on the right hand side then vanishes. Hence
	\begin{eqnarray}	\label{eq:dH1noise_lin}
	\dt {H_1^{noise}} = \frac12 \frac {g_{11}} {\sigma_{11}},
	\end{eqnarray}
where for notational symmetry $\sigma_1^2$ has been written as $\sigma_{11}$.
These, together with the information flow from $X_2$ to $X_1$ as we have 
obtained before\cite{LK05}\cite{Liang14},
%	\begin{eqnarray}	\label{eq:T21_lin}
	$T_{2\to1} = \frac{\sigma_{12}} {\sigma_{11}} a_{12},$
%	\end{eqnarray}
form the three constituents that account for the evolution of the marginal
entropy of $X_1$.

An observation about $\dt H_1^{noise} = g_{11} / (2\sigma_{11})$,
where 
	   $g_{11} = b_{11}^2 + b_{12}^2,$ 
is that it is always positive. That is to say, the noise always contributes
to increase the marginal entropy of $X_1$, conforming to our common sense.
In financial economics, this reflects the volatility of, say, a stock.
On the other hand, for a stationary series, i.e., when 
$d/dt=0$, the balance on the right hand side of Eq.~(\ref{eq:sigma}) 
requires that $2\sigma_{11} \sim g_{11}$. So this quantity is also related 
to the noise-to-signal ratio.

The above results need to be estimated if what we are given are just 
a pair of time series. 
That is to say, what we know is a single realization of some
unknown system, which, if known, can produce infinitely many realizations.
The problem now is turned into estimating
(\ref{eq:dH1star_lin}) and (\ref{eq:dH1noise_lin}) 
with the available statistics of the given time series.

% In this system-agnostic case, we are about infer the system from the only
% realization.

We use maximum likelihood estimation (e.g., \cite{Garthwaite}) 
to achieve the goal. 
The procedure follows precisely that of \cite{Liang14}, which
for easy reference we briefly summarize here.
As established before, a further assumption that 
$b_{12}=0$, and hence $g_{11}=b_{11}^2$, 
will much simplify the result, while in practice this is
quite reasonable. 

Suppose that the series are equal-distanced with a time stepsize $\delt$,
and let $N$ be the sample size.
Consider an interval $[n\delt, (n+1)\delt]$, and let the transition
probability function (pdf) be
$\rho(\ve X_{n+1} | \ve X_n; \veg\theta)$, where $\veg\theta$ stands for
the vector of parameters to be estimated. So the log likelihood is
	\begin{eqnarray*}
	\ell_N(\veg\theta) =
	\sum_{n=1}^N \log\rho(\ve X_{n+1} | \ve X_n; \veg\theta)
	+ \log\rho(\ve X_1).
	\end{eqnarray*}
As $N$ is usually large, the term $\rho(\ve X_1)$ can be dropped without
causing much error. The
transition pdf is, with the Euler-Bernstein approximation 
(see \cite{Liang14}),
	\begin{eqnarray*}
	\rho(\ve X_{n+1} = \ve x_{n+1} | \ve X_n = \ve x_n)
	= \frac 1 {[(2\pi)^2 \det(\vve B \vve B^T \delt) ]^{1/2}}
	  \times e^{-\frac12 (\ve x_{n+1} - \ve x_n - \ve F\delt)^T 
			     (\vve B\vve B^T\delt)^{-1} 
			     (\ve x_{n+1} - \ve x_n - \ve F\delt)},
	\end{eqnarray*}
where $\ve F = \ve f + \vve A\ve X$. This results in a log likelihood
functional
	\begin{eqnarray}
	\ell_N(\ve f, \vve A, \vve B)
	= \const - \frac N2 \log g_{11} g_{22}
	 -\frac\delt 2 \parenth{
		\frac 1 {g_{11}} \sum_{n=1}^N R_{1,n}^2 +
		\frac 1 {g_{22}} \sum_{n=1}^N R_{2,n}^2 
				},
	\end{eqnarray}
where 
	\begin{eqnarray}
	R_{i,n} = \dot X_{i,n} - (f_i + a_{i1} X_{1,n} + a_{i2} X_{2,n}),
	\qquad i=1,2,
	\end{eqnarray}
and $\dot X_{i,n}$ is the Euler forward differencing approximation
of $\dt {X_i}$: 
	\begin{eqnarray}
	\dot X_{i,n} = \frac {X_{i,n+k} - X_{i,n}} {k\delt},
	\end{eqnarray}
with $k\ge1$. Usually $k=1$ should be used to ensure accuracy, but in some
cases of deterministic chaos and the sampling is at the highest resolution,
one needs to choose $k=2$. Maximizing $\ell_N$,
we find\cite{Liang14} that the maximizer 
	$(\hat f_1, \hat a_{11}, \hat a_{12})$ 
satisfies the following algebraic equation:
	\begin{eqnarray}
	\matt 1                 {\overline{X_1}}   {\overline{X_2}}
	      {\overline{X_1}}  {\overline{X_1^2}} {\overline{X_1X_2}}  
	      {\overline{X_2}}  {\overline{X_1X_2}} {\overline{X_2^2}}  
	\vectt {\hat f_1} {\hat a_{11}} {\hat a_{12}}
	=
	\vectt {\overline{\dot X_1}} {\overline{X_1\dot X_1}}
	       {\overline{X_2 \dot X_1}},
	\end{eqnarray}
where the overline signifies sample mean. 
After some manipulations (see \cite{Liang14}), this yields the MLE estimators:
	\begin{eqnarray}
	&&\hat a_{11} = \frac{C_{22}C_{1,d1} - C_{12}C_{2,d1}} {\det\vve C},\\
	&&\hat a_{12} = \frac{-C_{12}C_{1,d1} + C_{11}C_{2,d1}} {\det\vve C},\\
	&&\hat g_{11} = \frac{Q_{N,1} \Delta t} N, \\
 	&& \hat f_1 = \bar {\dot X_1} - \hat a_{11} \bar X_1 
				      - \hat a_{12} \bar X_2,
	\end{eqnarray}
where 
	\begin{eqnarray}
	&&C_{ij} = \overline{(X_i - \bar X_i) (X_j - \bar X_j)},\\
	&&C_{i,dj} = \overline{(X_i - \bar X_i) (\dot X_j - \overline {\dot X_j})},
	\end{eqnarray}
are the sample covariances, and
	\begin{eqnarray}
	Q_{N,1} 
	&=& \sum_{n=1}^N \bracket{\dot X_{1,n} - 
	    (\hat f_1 + \hat a_{11} X_{1,n} + \hat a_{12} X_{2,n})}^2\cr
	&=& \sum_{n=1}^N \bracket{
	    (\dot X_{1,n} - \overline{\dot X_{1,n}})
	    - \hat a_{11} (X_{1,n} - \bar X_1) 
	    - \hat a_{12} (X_{2,n} - \bar X_2) }^2  	\cr
	&=& N (C_{d1,d1} + \hat a_{11}^2 C_{11} + \hat a_{12}^2 C_{22}
	    - 2\hat a_{11} C_{d1,1} - 2\hat a_{12} C_{d1,2}
	    + 2\hat a_{11} \hat a_{12} C_{12}).
	\end{eqnarray}

On the other hand, the population covariance matrix $\vveg\Sigma$ can
be rather accurately estimated by the sample covariance matrix $\vve C$. So 
(\ref{eq:dH1star_lin})-(\ref{eq:dH1noise_lin}) become
	\begin{eqnarray}
	&& \dt {H_1^*} = \frac{C_{22}C_{1,d1} - C_{12}C_{2,d1}} 
			   {C_{11}C_{22} - C_{12}^2},		\\
	&& \dt {H_1^{noise}} = \frac \delt {2 N C_{11}}	
				Q_{N,1}	 	
%	          \sum_{n=1}^N \bracket{\dot X_{1,n} - 
%			(\hat a_{11} X_{1,n} + \hat a_{12} X_{2,n})}^2,\\
	% &&T_{2\to1} = \frac {C_{11}C_{12}C_{2,d1} - C_{12}^2C_{1,d1}} 
	%		  {C_{11}^2 C_{22} - C_{11} C_{12}^2}.
	\end{eqnarray}
As that in Ref.~\cite{Liang14} with $T_{2\to1}$,
here $\dt {H_1^*}$ and $\dt H_1^{noise}$ should bear a hat, 
since they are the corresponding estimators. We abuse the notation a little bit
to avoid notational complexity; from now on they should be understood as
their respective estimators. 
With these the normalizer is
	\begin{eqnarray}
	Z = \abs{T_{2\to1}} + \abs{\dt {H_1^*}} + \abs{\dt {H_1^{noise}}},
	\end{eqnarray}
and hence we have the relative information flow from $X_2$ to $X_1$:
	\begin{eqnarray}	\label{eq:tau21}
	\tau_{2\to1} = \frac {T_{2\to1}} Z.
	\end{eqnarray}

\section{The autoregressive example revisited}	\label{sect:validation}
Back to the autoregressive process exemplified in the beginning. When
$\alpha=0$, $\beta=0$, the computed relative information flow rates are:
	\begin{eqnarray*}
	 \tau_{2\to1} = -0.0016\%, \qquad
	 \tau_{1\to2} =  0.0018\%.
	\end{eqnarray*}
Clearly both are negligible in comparison to the contributions from the
other processes in their respective series. This is in agreement with what
one would conclude based on the absolute information flow computation and
statistical testing. For the case $\alpha=\beta=0.01$, in which one may
encounter difficulty due to the ambiguous small numbers, the computed
relative information flow rates are:
	\begin{eqnarray*}
	 \tau_{2\to1} = 0.018\%, \qquad
	 \tau_{1\to2} =  0.015\%.
	\end{eqnarray*}
Again they are essentially negligible, just as one would expect.
	
On the other hand, when $\alpha=0.5$, $\beta=0$,
	\begin{eqnarray*}
	 \tau_{2\to1} = 17\%, \qquad
	 \tau_{1\to2} = -0.03\%.
	\end{eqnarray*}
To $X_1$, the influence from $X_2$ is large, contributing to more than 
1/6 of the total entropy change. In contrast, the influence from
$X_1$ to $X_2$ is negligible.

It should be pointed out that the relative information flow, say,
$\tau_{2\to1}$, makes sense only with respect to $X_1$, since the comparison
is within the series itself. Here comes the following situation: 
For a two-way causal system with absolute information flows 
$T_{2\to1}$ and $T_{1\to2}$ of equal importance, 
their relative importances within their respective series 
could be quite different.
For example, 
	\begin{eqnarray*}
	&& X_1(n+1) = -0.5 X_1(n) + 0.9X_2(n) + 2e_1(n),\\
	&& X_2(n+1) = -0.2X_1(n) + 0.5X_2(n) + e_2(n).
	\end{eqnarray*}
where $e_1(n)$ and $e_2(n)$ are identical independent normals ($\sim$N(0,1)).
Initialize them with random values between [0,1] and generate
80000 data points on MATLAB. 
The computed information flow rates (in nats per iteration)
	$$|T_{2\to1}| = 0.13,\qquad |T_{1\to2}|=0.12,$$
which are almost the same. 
The relative information flows, however, are quite different:
	$$|\tau_{2\to1}| = 6.7\%,\qquad |\tau_{1\to2}| = 13\%.$$ 
In terms of relative contribution in their respective series, 
the former is way more below the latter.

Generally speaking, the above imbalance is a rule, not an exception,
reflecting the asymmetry of information flow. One may reasonably imagine
that, in some extreme situation, a flow might be dominant 
while its counterpart is negligible within their respective series, 
although the two are of the same order in absolute value.

\section{Applications}	\label{sect:applications}
\subsection{Re-examination of the ENSO-IOD relationship} \label{sect:climate}

El Ni\~no, also known as El Ni\~no-Southern Oscillation, or ENSO for short, 
is a long known and extensively studied
climate mode in the tropical Pacific Ocean due to 
its relation to the global disasters like 
the droughts in Southeast Asia, Southern Africa, and northern Australia, 
the floods in Ecuador, the increasing number of Typhoons, 
the death of birds and dolphins in Peru, 
and the famine and epidemic diseases in far-flung regions of the
world\cite{ENSO}. 
	% and lasting nine months to two years.
A correct forecast of an \ENSO (or its cold counterpart, La Ni\~na) 
a few months earlier will not only help issue in-advance warnings of
potential disastrous impacts, but also make the subsequent 
seasonal weather forecasting much easier.
However, this aperiodic leading mode in the tropical Pacific 
seems to be extraordinarily difficult to predict.
A good example is the latest ``Super \ENSO'' or ``Monster \ENSO'',
which has been predicted to arrive in 2014 in a lot of portentous forecasts,
turns out to be a computer artifact.

% "One time scales from decades to months, fluctuations in ocean conditions
% present persistent challenges to climate scientists and weather
% forecasters. The latest example is the fadeout of portentous predictions of
% a super or monster El Nino warming of the tropical Pacific."

For more reliable predictions, it is imperative to clarify the source of
its uncertainty or unpredictability. 
In Ref.~\cite{Liang14}, we have presented an application 
of Eq.~(\ref{eq:T21_est}) to the relation study between \ENSO and the
Indian Ocean Dipole (IOD), another major climate mode in the Indian
Ocean\cite{Saji99},
and found that the Indian Ocean is a source of uncertainty that keeps
ENSO from being accurately predicted.
Since in that study there is no relative importance assessment, we do not
know whether the information flows, albeit significant, do weigh much in
the modal variabilities. We hence redo the computation using the relative 
information flow formula (\ref{eq:tau21}). 

We use for this study the same data as that used in \cite{Liang14},
which include the Ni\~no4 index series and the sea surface temperature (SST)
series downloaded from the NOAA ESRL Physical Sciences Division\cite{Nino4},
and the IOD index namely DMI series from the JAMSTEC site\cite{DMI}. 

\begin{figure}[h]
	\begin{center}
	\includegraphics[angle=0, width=0.6\textwidth]
		{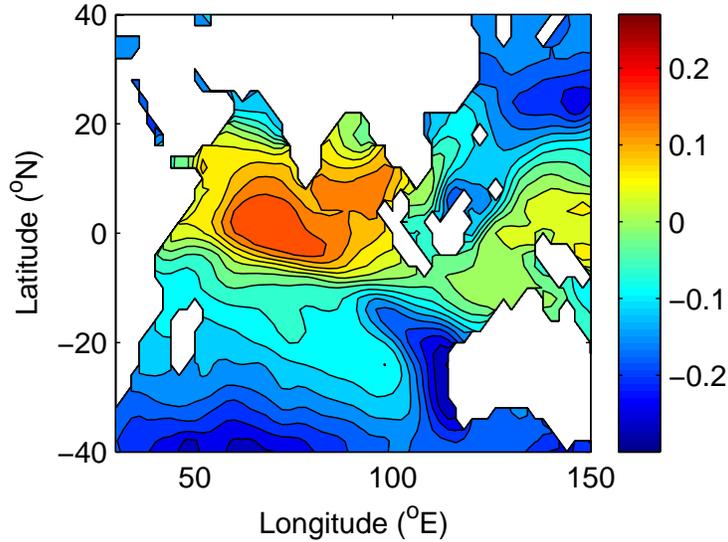}
	\caption{(Color online) Relative information flow from
	the Indian Ocean SST to Ni\~no4,
	$\tau_{IO\to ENSO}$.
	 \protect{\label{fig:Indian}}}
	\end{center}
	\end{figure}

Shown in Fig.~\ref{fig:Indian} is the relative information flow rate from
the Indian Ocean SST to El Ni\~no, $\tau_{IO\to ENSO}$. From it one can see
that the information flow accounts for more than 10\% of the uncertainties
of Ni\~no4, the maximum reaching 27\%. This number is very large.
Besides, all the values are positive, indicating that the Indian Ocean SST
functions to make \ENSO more uncertain. No wonder recently
researchers find that assimilation of the Indian Ocean data helps
the prediction of \ENSO (e.g., \cite{Chen08}), although traditionally the
Indian Ocean is mostly ignored in \ENSO modeling.

Besides the relative importance we have just obtained,
Fig.~\ref{fig:Indian} also reveals some difference in structure from its
counterpart, i.e., the Fig.~5b of \cite{Liang14}.\footnote
    {(Note: There is a typo in the caption of the Fig.~5 of 
	\cite{Liang14}. The units should be nats/year,  
	 not nats/month.)} 
A conspicuous difference is that now there are
clearly two centers, residing on either side of Indian. Note this structure
is different from the traditional dipolar pattern as one would expect;
here both centers are positive. 
This means that the Northern Indian Ocean SST anomalies,
both the positive phase and negative phase, 
as an integral entity influence the \ENSO variabilities, and, in
particular, make \ENSO more unpredictable.
The dipolar structure implies that most probably this entity is IOD, 
not others like IOBM (Indian Ocean Basin Mode; see \cite{IOBM}). 

\begin{figure}[h]
	\begin{center}
	\includegraphics[angle=0, width=1\textwidth]
		{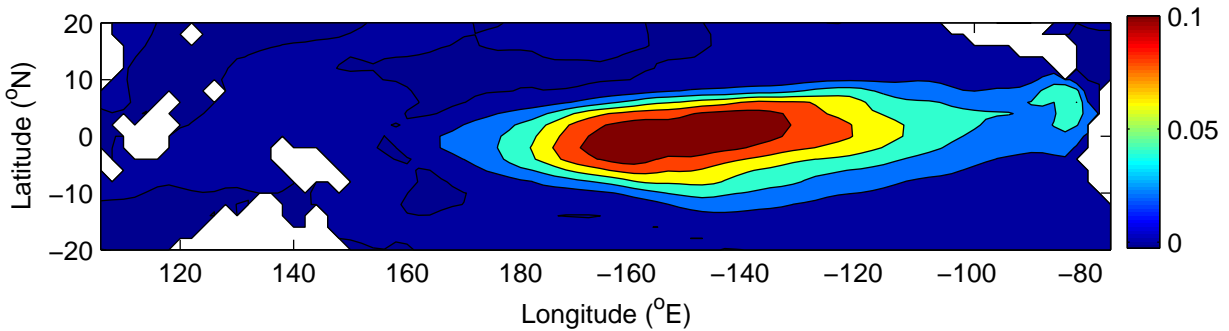}
	\caption{(Color online) Relative information flow from the
	IOD index to the tropical Pacific SST, $\tau_{IOD\to Pacific}$.
	 \protect{\label{fig:Pacific}}}
	\end{center}
	\end{figure}

To see more about this, 
we look at the information flow from the index DMI to the tropic 
Pacific SST. The absolute rates are referred to the
Fig.~4a of \cite{Liang14}; shown in Fig.~\ref{fig:Pacific} are
$\tau_{DMI\to Pacific}$. Indeed the computed flow rates are
significant, and all are positive. 
The largest $\tau$, which occupies a large swathe of the equatorial region
between $\rm 175^oW$  through $\rm 135^oW$, reaches 10\%. 
Moreover, the structure reminds one of the \ENSO pattern.
It is generally the same as that in its counter part, i.e., 
Fig.~4a of \cite{Liang14}, save for two changes: (1) the maximum center moves
westward; (2) the small center of a secondary maximum near $\rm 125^oE$ at the
equator disappears. This clear El Ni\~no-like structure attests to the above
conjecture that IOD is indeed a major source of uncertainty for the \ENSO
forecast.

We have also computed the relative information flows from \ENSO to the
Indian Ocean SST, and that from the Pacific SST to the IOD, using the same
datasets. The results are also significant, though only approximately half
as shown above.

\subsection{A financial example}	\label{sect:finance}

We now look at the causal relations between several financial time series. 
Here it is not our intention to conduct an financial
economics research or study the market dynamics from an econophysical 
point of view; our purpose is to demonstrate a brief application of the
aforementioned formalism for time series analysis.
Nonetheless, this topic is indeed of interest to both physicists and
economists in the field of macroscopic econophysics; see, for example,
\cite{Kantz}.

%
% "There has been a recent but growing interest of physicists in the dynamics
% of financial
% markets\cite{Kantz}\cite{Bouchaud}\cite{Lux}\cite{Ghashghaie}\cite{Mantegna}.
% This might have been stimulated by the discovery of some strong analogies
% between speculative markets and some well known physical phenomena and
% concepts, as for instance spin system[8], turbulence[3], universality[7],
% self-organized criticality[5,9], and complexity[6], almost all of which can be
% associated with the statistical mechanics branch of physics. With growing
% recognition of this new field of interest the term econophysics was being
% conined."
%
% "Today, one can divide the research activities within econophysics roughly
% in two areas: the microscopic approach investigates the financial market
% dynamics from the pt of view of the single agents, with the long-term
% target of being able to reproduce the complex "macroscopic" behavior of the
% financial markets starting from microscopic equations [5,8]. To thoroughly
% analyze and describe the statistical properties of that macroscropic
% behavior is exactly what constitutes the second branck of
% econophysics[10-14]. .... info flow analysis is this..."

We pick nine stocks in the United States and download their daily prices
from ${YAHOO!} \atop {\rm {finance}}$. These stocks are: 
MSFT (Microsoft Corporation), 
AAPL (Apple Inc.), 
IBM (International Business Machines Corporation), 
INTC (Intel Corporation), 
GE (General Electric Company), 
WMT (Wal-Mart Stores Inc.),
XOM (Exxon Mobil Corporation), 
CVS (CVS Health Corporation), 
F (Ford Motor Corporation).
Among these are high-tech companies (MSFT, AAPL, IBM, INTC),
retail trade companies [e.g., the drugstore chains (CVS) and discount
stores (WMT)],
automotive industry (F),
oil and gas industry (XOM),
and the multinational conglomerate corporation GE 
which operates through the segments of energy, technology 
infrastructure, capital finance, etc.
Here by ``daily'' we mean on a trading day basis, 
excluding, say, holidays and weekends.
Since stock prices are generally nonstationary, we check the series 
of daily return, i.e., 
	$$R(t) = [P(t + \Delta t) - P(t)] / P(t),$$
or log-return 
	$$r(t) = \ln P(t+\Delta t) - \ln P(t),$$
where $P(t)$ are the adjusted closing prices in the yahoo spreadsheet,
and $\Delta t$ is one trading day.
Following most people we use the series of log-returns $r$ for our purpose.
In fact, return and log-return series are approximately equivalent, 
particularly in the 
high-frequency regime, as indicated by \cite{Mantegna}.
Since the most recent stock MSFT started on March 13, 1986, all the series
are chosen from that date through December 26, 2014,
	when we started to examine these series.
This amounts to 7260 data points, and hence 7259 points for the
log-return series.

Using Eq.~(\ref{eq:T21_est}), we compute the information flows between the 
nine stocks and form a matrix of flow rates; see Table~\ref{tab:stocks}.
A flow direction is represented with the matrix indices; more specifically,
it is from the row index to the column index. 
For example, listed at the location (2,4) is 
$T_{2\to4}$, i.e., $T_{AAPL\to INTC}$, the flow rate from Apple to Intel,
while (4,2) stores the rate of the reverse flow, $T_{INTC\to AAPL}$.
Also listed in the table are the respective confidence
intervals at the 90\% level. 

From Table~{\ref{tab:stocks}}, most of the information flow rates are 
significant at the 90\% level, as highlighted. Their values vary from 
4 to 22 (units: $10^{-3}$~nats/day; same below in this section).
The maximum is $|T_{IBM\to XOM}| = 22$, and second to it are
$|T_{WMT\to CVS}|$ and $|T_{CVS\to GE}|$, both being $21$.
The influence of IBM to Exxon is not a surprise, considering the dependence
of the oil industry on high-tech equipments.
The mutual causality between the retail stores WMT and CVS are also
understandable. The information flow from CVS to GE could be through
the sales of GE products; after all, GE makes household appliances. 
For the rest in the table, they can be summarized from 
the following two aspects.
	\begin{itemize} 
	\item[(1)] Companies as sources.\\
	Look at the table row by row. 
	Perhaps the most conspicuous feature is that
	the whole CVS row is significant. Next to it is XOM, with only
	three entries insignificant. 
	      % AAPL, INTC, GE, and WMT rows are the third.
	That is to say, CVS has been found causal to all other stocks,
	though the causality magnitudes are yet to be assessed (see below).
	This does make sense. As a chain of convenience stores, CVS
	connects most of the general consumers and commodities and hence the
	corresponding industries. 
	For XOM, it is also understandable that why it makes a source  of
	causality. Oil or gas is for sure one of the most fundamental
	component in the American economy.  
	\item[(2)] Companies as recipients.\\
	Examining column by column, the most outstanding stock is again
	CVS, with only one entry insignificant. 
	That is to say, CVS is influenced by all other stocks
	except XOM. Following CVS is XOM, WMT, and INTC. 
	The IBM and MSFT columns form the third tier.
	\end{itemize}

A few words regarding the stock F.
As a cause to other stocks (though causality maybe tiny), 
XOM has not been identified
to be causal to F. In fact, F has not been found causal to XOM, either.
This is a little surprising; the reason(s) can be found only after a
careful analysis of Ford, which is beyond the scope here. (In fact,
computation does reveal information flows between XOM and Toyota.)
Interestingly, $|T_{F\to WMT}| > |T_{F\to CVS}|$. 
This is easy to understand, as we rely on our motor 
vehicles to shop at Wal-Mart, while CVS stores
could be just somewhere in the neighborhood!

\begin{table}[h]
\begin{center}
\caption{The rates of absolute information flow between the 9 chosen stocks 
	 (in $10^{-3}$ nats per trading day). 
	 At each entry the direction is from the 
	 row index to the column index of the matrix. Also listed are the
	 standard errors at a 90\% significance level, and highlighted
	 are the significant flows.
	 \protect{\label{tab:stocks}}}
\begin{tabular}{cccccccccc}
\hline
\hline
 \ & MSFT & AAPL & IBM & INTC & GE & WMT & XOM & CVS & F \\
\hline
MSFT & / &  5$\pm$7 & -3$\pm$8 & \highlight{-12$\pm$11} & 1$\pm$8 &
	-1$\pm$6 & -\highlight{12$\pm$6} & \highlight{10$\pm$4} & 3$\pm$5 \\
AAPL & -2$\pm$7 & / & \highlight{-11$\pm$7} & -2$\pm$9 & \highlight{-7$\pm$6} 
	& -4$\pm$5 & \highlight{-11$\pm$4} & \highlight{4$\pm$3} & -2$\pm$4 \\
IBM  & 0$\pm$8 & 5$\pm$7& / & -9$\pm$9 & -8$\pm$9 &\highlight{-11$\pm$6}
	& \highlight{-22$\pm$7 } & \highlight{6$\pm$4} & 1$\pm$6 \\
INTC & \highlight{16$\pm$11} & \highlight{10$\pm$9} & -7$\pm$9 & / & 0$\pm$8 
	& -2$\pm$6 & \highlight{-12$\pm$5} & \highlight{11$\pm$4} & 3$\pm$6 \\
GE   & 2$\pm$8 & -3$\pm$6& \highlight{-13$\pm$9} & \highlight{-16$\pm$8}
 	& / & \highlight{-10$\pm$9} & -6$\pm$9 & \highlight{14$\pm$6} 
	& 6$\pm$9 \\
WMT  & \highlight{10$\pm$6} & \highlight{7$\pm$5} & 4$\pm$6 & -5$\pm$6 
	& 6$\pm$9 & /  & 0$\pm$6 & \highlight{21$\pm$7} & \highlight{9$\pm$5}\\
XOM  & \highlight{-10$\pm$6} & -3$\pm$4 & \highlight{-14$\pm$7} 
	& \highlight{-13$\pm$5} & \highlight{-15$\pm$9} & \highlight{-17$\pm$6}
	& / & 4$\pm$5 & -1$\pm$6 \\
CVS  & \highlight{-9$\pm$4} & \highlight{-5$\pm$3} & \highlight{-12$\pm$4} 
	& \highlight{-11$\pm$4} & \highlight{-21$\pm$6} 
	& \highlight{-17$\pm$7} & \highlight{-17$\pm$5} & /  
	& \highlight{-7$\pm$4} \\
F    & 0$\pm$5 & 0$\pm$4& 0$\pm$6 & \highlight{-10$\pm$6} & 6$\pm$9 
	& \highlight{-13$\pm$5} & 1$\pm$6 & \highlight{6$\pm$4} & /  \\
\hline
\end{tabular}
\end{center}
\end{table}

The above significant absolute information flows, large or small, still
need to assessed regarding their respective relative importances before
any conclusion of causality is reached.
Using Eq.~(\ref{eq:tau21}), we compute the relative information flow rates
(in percentage) and tabulate them in Table~\ref{tab:stocks_norm}. 
The matrix entries are arranged in the same way as above. For clarity, 
those above or equal to 1\% are highlighted.
In contrast to Table~\ref{tab:stocks}, we see only a few of the information
flows account for more than 1\% of their respective fluctuations. 
These include
	$|\tau_{IBM\to XOM}|$,
	$|\tau_{CVS\to GE}|$,
	$|\tau_{WMT\to CVS}|$,
	$|\tau_{CVS\to WMT}|$,
	$|\tau_{CVS\to XOM}|$,
	$|\tau_{XOM\to WMT}|$,
	$|\tau_{INTC\to MSFT}|$,
	$|\tau_{GE\to INTC}|$, and
	$|\tau_{XOM\to GE}|$, the first three being the largest.
This echos what we have introduced in the beginning: though significant,
some information flows may be negligible in their own marginal entropy balances.

\begin{table}[h]
\begin{center}
\caption{As Table~{\ref{tab:stocks}}, 
	 but for relative information flow (in percentage). 
	 \protect{\label{tab:stocks_norm}}}
%\begin{tabular}{cccccccccc}
%\hline
%\hline
% \ & MSFT & AAPL & IBM & INTC & GE & WMT & XOM & CVS & F \\
%\hline
%% MSFT & / &  0.0&   0.0&   0.0&   0.0&   0.0&   0.0&   0.0&   0.0\\
%MSFT & / & 0.0\%&   0.0\%&   0.0\%&   0.0\%&   0.0\%&   0.0\%&   0.0\%& 0.0\%\\
%AAPL & 0.0\%& /&   0.0\%&   0.0\%&   0.0\%&   0.0\%&   0.0\%&   0.0\%& 0.0\%\\
%IBM  & 0.0\%& 0.0\%& / &   0.0\%&   3.1\%&   0.0\%&   0.0\%&   0.0\%& 0.0\%\\
%INTC & 0.0\%& 0.0\%& 0.0\% &  0.0\%&  0.0\%&   0.0\%&   0.0\%&   0.0\%& 0.0\%\\
%GE  & 0.0\%& 0.0\%& -0.1\% &  0.0\%&  / &   0.0\%&   0.0\%&   0.0\%& 0.0\%\\
%WMT & 0.0\%& 0.0\%&  0.0\% &  0.0\%&  0.0\%&   /&   0.0\%&   0.0\%& 0.0\%\\
%XOM & 0.0\%& 0.0\%&  0.0\% &  0.0\%&  0.0\%&  0.0\%& /&   0.0\%& 0.0\%\\
%CVS & 0.0\%& 0.0\%&  0.0\% &  0.0\%&  0.0\%&  0.0\%& 0.0\%& /& 0.0\%\\
%F & 0.0\%& 0.0\%&  0.0\% &  0.0\%&  0.0\%&  0.0\%& 0.0\%& 0.0\%& /\\
%\hline
%\end{tabular}
%
\begin{tabular}{cccccccccc}
\hline
\hline
 \ & MSFT & AAPL & IBM & INTC & GE & WMT & XOM & CVS & F \\
\hline
MSFT & / & 0.3 &   -0.2 &  -0.8 &   0.0 &   0.0 &  -0.7 &   0.6 & 0.2 \\
AAPL &-0.1 & /&  -0.7 &  -0.1 &  -0.5 &  -0.2 &  -0.7 &   0.2 &-0.1 \\
IBM  & 0.0 & 0.3 & / &  -0.6 &  -0.5 &  -0.7 &\highlight{-1.3} &   0.4 & 0.1 \\
INTC &\highlight{1.0} & 0.7 &-0.4  &  / &  0.0 &  -0.1 &  -0.7 &   0.7 & 0.2 \\
GE  & 0.1 &-0.2 & -0.8  & \highlight{1.0} &  / & -0.6 &  -0.3 &   0.9 & 0.4 \\
WMT & 0.6 & 0.4 &  0.2  & -0.3 &  0.4 &   /&   0.0 &\highlight{1.3} & 0.6 \\
XOM &-0.6 &-0.2 & -0.9  & -0.9 &\highlight{-1.0} &\highlight{-1.1} & /&   0.3 &-0.1 \\
CVS &-0.6 &-0.3 & -0.8  & -0.7 &\highlight{-1.3} &\highlight{-1.1} &\highlight{-1.1} & /& -0.5 \\
F & 0.0 & 0.0 &  0.0  & -0.6 &  0.4 & -0.8 & 0.0 & 0.4 & /\\
\hline
\end{tabular}
\end{center}
\end{table}

\begin{figure}[h]
	\begin{center}
	\includegraphics[angle=0, width=0.75\textwidth]
		{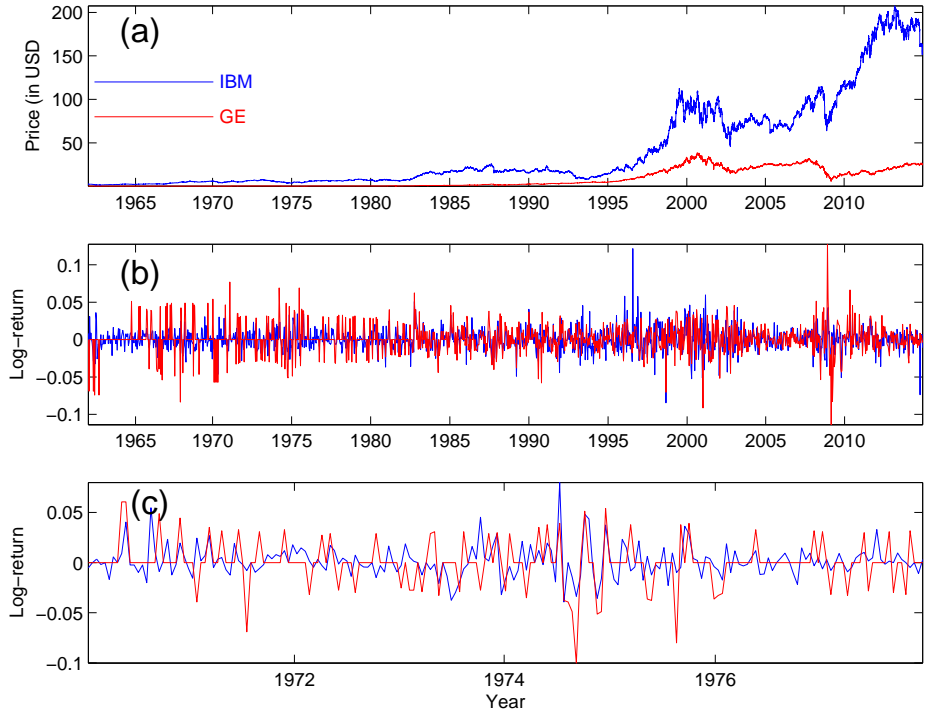}
	\caption{(Color online) 
	(a) Historic prices of IBM and GE (in US
	dollars). (b) Log-returns of the IBM and GE stocks.
	(c) A close-up of (b).
	 \protect{\label{fig:ibm_ge}}}
	\end{center}
	\end{figure}

It should be noted that the causal relations generally change with time. 
If the series are long enough, we may look at how these information flows 
may vary from period to period. Pick the pair (IBM,~GE) as an example.
For the duration (March 1986 through present) 
considered above, $T_{GE\to IBM} = -13 \pm 9$,
while $T_{IBM\to GE}$ is not significant. Neither $\tau_{GE\to IBM}$ 
nor $\tau_{IBM\to GE}$ reaches 1\%.
Since from the yahoo site both GE and IBM can be dated back to 
January 2, 1962, we can extend the time series a lot up to
13338 data points. Shown in Fig.~\ref{fig:ibm_ge}a are the series
of their historic prices, and in Fig.~\ref{fig:ibm_ge}b and \ref{fig:ibm_ge}
c are the corresponding log-returns.

Computation of information flows with the whole series (13338 points)
results in $\tau_{IBM\to GE} = 1.6\%$ and $\tau_{GE\to IBM} = -0.5\%$,
and $T_{IBM\to GE} = (27 \pm 6) \times 10^{-3}$,
    $T_{GE\to IBM} = (7 \pm 6) \times 10^{-3}$~nats/day, 
both being significant at the 90\% level. This is very different from what
are shown in Tables~\ref{tab:stocks} and \ref{tab:stocks_norm}, with the
causal structure changed from a weak two-way causality to a stronger
and more or less one-way causality. Since in the above only the data of
the recent 30 years are used, we expect that in the early years this causal
structure could be much enhanced. Choose the first 7000 points (
from January 1962 through November 1989), 
the computed relative information flow rates are:
	\begin{eqnarray*}
	\tau_{IBM\to GE} = 3.1\%, \ \ \ \tau_{GE\to IBM} = -0.2\%;  \\
	T_{IBM\to GE} = 54\pm8, \  \ \ \ T_{GE\to IBM} = 3\pm8,
	\end{eqnarray*}
where the units for the latter pair are in $10^{-3}$~nats/day, same below
in this section. Further narrow down the period to 2250-3250 (corresponding 
to the period 1971-1975), then
	\begin{eqnarray*}
	\tau_{IBM\to GE} = 5.7\%, \ \ \ \tau_{GE\to IBM} = -0.99\%;  \\
	T_{IBM\to GE} = 101\pm21, \  \ \ \ T_{GE\to IBM} = 14\pm21,
	\end{eqnarray*}
attaining the maximum of $\tau_{IBM\to GE}$, in contrast to the
insignificant flow in Table~\ref{tab:stocks}. 
Obviously, during this period, 
the causality can be viewed as one-way, i.e., from IBM to GE.
And the relative flow makes more than 5\%, much larger than those in 
Table~\ref{tab:stocks_norm}.

The above remarkable causal structure for that particular period actually
can trace its reason back in the history of GE\cite{GEhistory}. 
There is such a period in 1960's when ``Seven Dwarfs'' 
(Burroughs, Sperry Rand, Control Data,
Honeywell, General Electric, RCA and NCR) competed with IBM the giant 
for computer business, and, particularly, to build mainframes. 
In 1965, GE had only a 3.7-percent market share of the
industry, though it was then dubbed as the ``King of the Dwarfs'',
while IBM had 65.3\% share. 
Historically GE was once the largest computer user outside 
the US Federal Government; it got into computer manufacturing 
to avoid dependency on others. And, indeed, throughout the 60s, the
causalities between GE and IBM are not significant. Then, why, as time
entered 70s, was the information flow from IBM to GE suddenly increased
to its highest level?
It turned out that GE sold its computer division to Honeywell in 1970;
in the following years (starting from 1971), it relied much on the 
IBM products.
This GE computer era, which has almost gone to oblivion,
	% is ironically missing from the company's website, 
does substantiate the existence of a causation between GE and
IBM, and, to be more precise, an essentially one-way causation from IBM to
GE. In this sense, our formalism is well validated.

%"If you go to the GE website and attempt to find the history of the
%company's foray into computers you simply won't find it. Weirder still is
%that the entire episode is completely missing from the company's timeline
%at www.ge.com/ibhist2.htm."

% Also compute noise contributions:
% volatility for each stock

\section{Conclusions}	\label{sect:summary}

\begin{figure}[h]
	\begin{center}
	\includegraphics[angle=0, width=0.65\textwidth]
		{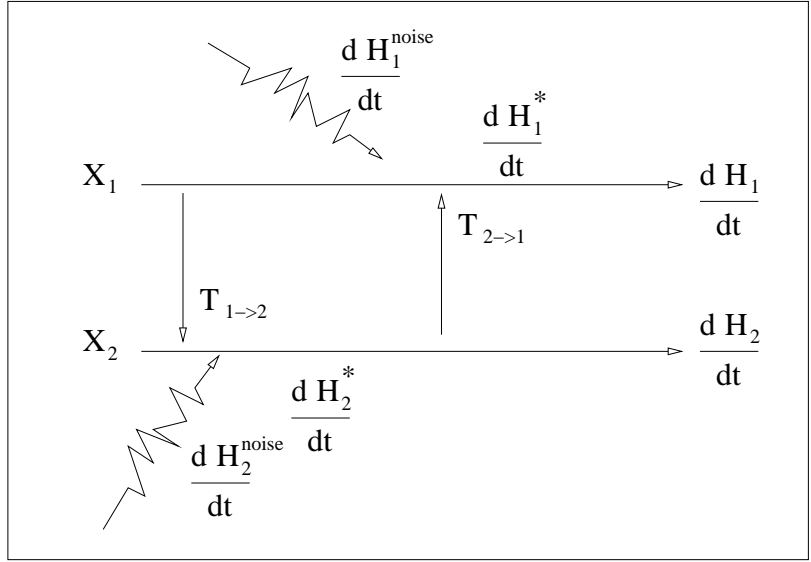}
	\caption{A schematic of the marginal entropy evolutions and
	        information flows in the system of ($X_1$,~$X_2$).
	 \protect{\label{fig:schem_info_flow}}}
	\end{center}
	\end{figure}

To assess the importance of a flow of information from a series, say,
$\{X_2(n)\}$, to another, say, $\{X_1(n)\}$, it needs to be normalized. 
The normalization cannot follow a way as that in computing
a correlation coefficient, 
since there is no such a theorem like the Cauchy-Schwarz inequality for it
to base. Getting down to the fundamentals, we were able to distinguish three
types of mechanisms that contribute to the evolution of the marginal
entropy of $X_1$, $H_1$, 
	\begin{itemize}
	\item $\dt {H_1^*}$: the contribution from $X_1$ its own; 
	\item $T_{2\to1}$: the information flow from $X_2$;
	\item $\dt {H_1^{noise}}$: the contribution from noise.
	\end{itemize}
Similarly there are three such quantities for $X_2$, as
schematized in Fig.~\ref{fig:schem_info_flow}. 
We hence proposed that the normalization can be fulfilled as follows:
	\begin{eqnarray*}
	\tau_{2\to1} &=&  \frac{T_{2\to1}} 
	     {\abs{\dt {H_1^*}} + \abs{\dt {H_1^{noise}}} + \abs{T_{2\to1}}},
						\\
	\tau_{1\to2} &=&  \frac{T_{1\to2}} 
	     {\abs{\dt {H_2^*}} + \abs{\dt {H_2^{noise}}} + \abs{T_{1\to2}}}.
	\end{eqnarray*}
Obviously, a normalized flow tells its importance relative to other
mechanisms within its own series. In other words, the two flows are
normalized differently, echoing the property of asymmetry which makes
information flow analysis distinctly different from those such as
correlation analysis or mutual information analysis.

The above normalizer can be accurately obtained in the framework of a
dynamical system. When only two equi-distanced series, say, 
$X_1$ and $X_2$, are given, with a linear model its constituents 
can be estimated as follows, in a nutshell,
	\begin{eqnarray}
	&& \dt {H_1^*} = p,		\\
	&& \dt {H_1^{noise}} = \frac \delt {2 C_{11}}	
		\bracket{C_{d1,d1} + p^2 C_{11} + q^2 C_{22}
			-2p C_{d1,1} - 2q C_{d1,2} + 2pq C_{12} },	\\
	&&T_{2\to1} = \frac {C_{11}C_{12}C_{2,d1} - C_{12}^2C_{1,d1}} 
			  {C_{11}^2 C_{22} - C_{11} C_{12}^2},
	\end{eqnarray}
where $\delt$ is the time stepsize, 
	\begin{eqnarray*}
	&& p = \frac{C_{22}C_{1,d1} - C_{12}C_{2,d1}} 
			   {C_{11}C_{22} - C_{12}^2},		\\
	&& q = \frac{-C_{12}C_{1,d1} + C_{11}C_{2,d1}} 
			   {C_{11}C_{22} - C_{12}^2},		\\
	\end{eqnarray*}
$C_{ij}$ the sample covariance between $X_i$ and $X_j$,
$C_{i,dj}$ the sample covariance between $X_i$ and $\dot X_j$, and
	$$\dot X_{i,n} = \frac {X_{i,n+k} - X_{i,n}} {k\delt}$$
($k=1$; but for chaotic series sampled at high resolution, $k=2$ may be
needed).

%	\begin{eqnarray*}
%	&& \dt {H_1^*} = \frac{C_{22}C_{1,d1} - C_{12}C_{2,d1}} 
%			   {C_{11}C_{22} - C_{12}^2},		\\
%	&& \dt {H_1^{noise}} = \frac \delt {2 N C_{11}}
%	          \sum_{n=1}^N \bracket{\dot X_{1,n} - 
%		(\hat f_1 + \hat a_{11} X_{1,n} + \hat a_{12} X_{2,n})}^2,\\
%	&&T_{2\to1} = \frac {C_{11}C_{12}C_{2,d1} - C_{12}^2C_{1,d1}} 
%			  {C_{11}^2 C_{22} - C_{11} C_{12}^2},
%	\end{eqnarray*}

It should be noted that a relative information flow is for the comparison
purpose within its own series. The two reverse flows between two series can
only be compared in terms of absolute value, since they belong to different
series. In this sense, absolute and relative information flows should be
examined simultaneously. This is clarified in the schematic diagram in
Fig.~\ref{fig:schem_info_flow}, and has been exemplified in the validations
with two autoregressive processes.
It is quite normal that two identical information flows may differ a lot
in relative importance with respect to their own series, as testified
in our realistic applications.
In some extreme situation, a pair of equal flows may find 
one dominant but another negligible in their respective entropy balances. 

Partly for demonstration and partly for verification, we have presented two
applications. The first is a re-examination of the climate science problem
previously studied in Ref.~\cite{Liang14}. Considering the fadeout of the 
recent portentous predictions of a ``super'' or ``monster'' El Ni\~no, 
we have particularly focused on the predictability of El Ni\~no.
Our result reconfirmed that the Indian Ocean 
SST is a source of uncertainty to the \ENSO prediction. We further
clarified that the information flow from the Indian Ocean is mainly through
the Indian Ocean Dipole (IOD).

Another realistic problem we have examined regards the causation between a
few randomly picked American stocks. 
% These stocks were originally picked
% simply for the verification of our formalism and 
% the resulting tight and concise information flow formula. 
It is shown that many flows (and hence causalities),
though significant at a 90\% level, their respective importances relative
to other mechanisms are mostly negligible. The resulting matrices of absolute
and relative information flows provide us a pattern of causality mostly
understandable using our common sense. 
For example, Ford has a larger influence on Wal-Mart than on CVS because
people rely on motor vehicles to shop at Wal-Mart, while CVS could be just
somewhere within a walking distance.
A particularly interesting case is
that we have identified a strong one-way causality from IBM to GE during
the early stage of these companies. This has revealed to us the story of 
``Seven Dwarfs'' competing IBM the giant for computer market. In an era
when this story has almost gone to oblivion (one even cannot find it from GE's
website), and GE may have left us an impression that it never built any
computers, let alone a series of mainframes, this finding is indeed
remarkable.

\vskip 0.5cm 

\noindent
{\bf Acknowledgments.}
This study was partially supported by Jiangsu Provincial Government through
the ``Specially-Appointed Professor Program'' 
(Jiangsu Chair Professorship) to XSL, and by
the National Science Foundation of China (NSFC) under Grant No.~41276032.


\begin{thebibliography}{99}

%%%%%%%%%%%%%%%%%%%%%%%%%%
% cite Tom Haine's paper
%%%%%%%%%%%%%%%%%%%%%%%%%%

\bibitem{causalmeasure} 
% examples of references that take information flow as causality measure
e.g., K. Hlav\'a$\rm\breve c$kov\'a-Schindler, M. Palu$\rm\breve s$, 
M. Vejmelka, J.  Bhattacharya, 
   % Causality detection based on information-theoretic approaches in
   % time series analysis. 
  {Physics Reports}, 441(1), 1-46 (2007);
J. Pearl, {\em Causality: Models, Reasoning, and Inference}. MIT Press, 
Cambridge, MA, 2nd edition (2009);
M. Lungarella, K. Ishiguro, Y. Kuniyoshi, N. Otsu,
   % Methods for quantifying the causal structgure of bivariate time
   % series.
{International Journal of Bifurcation and Chaos}, 17(3), 903-921 (2007). 


\bibitem{Granger} C. Granger, 
 % Investigating causal relations by econometric models and cross-spectral
 % methods.
 {\it Econometrica}, 37, 424 (1969).

\bibitem{Swinney} J.A. Vastano and H.L. Swinney, 
	Phys. Rev. Lett. {\bf 60}, 1773 (1988).

\bibitem{Schreiber} T. Schreiber, {Phys. Rev. Lett.} {\bf 85}, 461 (2000).

\bibitem{Barnett} L. Barnett, A.B. Barrett, and A.K. Seth, 
	{Phys. Rev. Lett.} {\bf 103}, 238701 (2009).

\bibitem{Runge} J. Runge, J. Heitzig, N. Marwan, 
	and J. Kurths, Phys. Rev. E 86, 061121 (2012).

\bibitem{JieSun} J. Sun and E. Bolt, Physica D, 267, 49-57 (2014).
%	Causation entropy identifies indirect influences, dominance of
%	neighbors, and anticipatory couplings.

%\bibitem{Smirnov} D. Smirnov,
%	% Spurious causalities with transfer entropy.
%	{Phys. Rev. E}, 87(4), 042917 (2013).

\bibitem{LK05} X.S. Liang, R. Kleeman, 
	{Phys. Rev. Lett.} {\bf 95}, 244101 (2005);
%	X.S. Liang, and R. Kleeman, {Physica D} (2007ab)...

\bibitem{Liang13} X.S. Liang,
	{Entropy} {\bf 15}, 327 (2013).

\bibitem{Liang08} X.S. Liang,
	{Phys. Rev. E} {\bf 78}, 031113 (2008).



\bibitem{Liang14} X.S. Liang,  Unraveling the cause-effect relation between
time series. {Phys. Rev. E} {\bf 90}, 052150 (2014).



%%%%%%%%%%% Estimation %%%%%%%%%%%%%
\bibitem{Garthwaite} P.H. Garthwaite, I.T. Jolliffe, and B. Jones,
	{\it Statistical Inference}. (Prentice Hall, Hertfordshire, UK
	1995).






%%%%%%%%%% ENSO-IOD %%%%%%%%%%%%
\bibitem{ENSO} e.g., M.A. Cane, {Science} {\bf 222}, 1189 (1983);
J.M. Wallace, E.M. Rasmusson, PP. Mitchell, V.E. Kousky, E.S. Sarachik, H.
von Storch, {\it J. Geophys. Res.} {\bf 103}, 14241 (1998).

\bibitem{Saji99} H.H. Saji, B.N. Goswami, P.N. Vinayachandran, 
	and T. Yamagata, {Nature} {\bf 401}, 360 (1999).

\bibitem{Nino4}
	http://www.esrl.noaa.gov/psd/. 
\bibitem{DMI}
	http://www.jamstec.go.jp/.



\bibitem{Chen08} D. Chen, M.A. Cane,
	{J. Comput. Phys.} {\bf 227}, 3625 (2008).





\bibitem{IOBM}
J.L. Yang, Q.Y. Liu, and Z. Liu: 
Linking observations of the Asian Monsoon to the Indian Ocean SST: 
Possible roles of Indian Ocean Basin Mode and Dipole Mode.
J. Clim. 2010, 23(21), 5889-5902.





% financial economics
\bibitem{Kantz} R. Marshinski and H. Kantz, Analysing the information flow
between financial time series: An improved estimator for transfer entropy.
     	{Eur. Phys. J. B} {\bf 30}, 275-281 (2002).

% \bibitem{Bouchaud}
% J.-P. Bouchaud, M. Potters, Theory of financial risks (Cambridge University
%	Press, Cambridge, 2000)
%
%% Self-organized criticality 
% \bibitem{Lux}
% T. Lux, M. Marchesi, Nature 397, 498 (1999)
% \bibitem{Sornette}
% D. Sornette, A. Johansen, J.-P. Bouchaud, J. Phys. I 6, 167 (1996)
%
%% Turbulence
% \bibitem{Ghashghaie}
% S. Ghashghaie, W. Breymann, J. Peinke, P. Talkner, Y. Dodge, Nature 381,
%	767 (1996)
%
%% Spin system
% D. Chowdhury, D. Stauffer, Eur. Phys. J. B 8, 477 (1999)
%
%% Universality
% V. Plerou, P. Gopikrishnan, B. Rosenow, L.A.N. Amaral, H.E. Stanley, Phys.
% Rev. Lett. 83, 1471 (1999)
%
% Complexity
\bibitem{Mantegna}
R. N. Mantegna, H.E. Stanley, {\it An Introduction to Econophysics. Correlations
and complexity in finance} (Cambridge University Press, Cambridge, 2000)


\bibitem{GEhistory}http://en.wikipedia.org/wiki/General\_Electric/

\end{thebibliography}
\end{document}